\documentclass[9pt,conference]{IEEEtran}
\IEEEoverridecommandlockouts
\usepackage{amsmath, bbm}
\usepackage{amssymb}
\usepackage{amsfonts}
\usepackage{amsthm}
\usepackage{algorithm}
\usepackage{algorithmic}
\usepackage{nicefrac}
\usepackage{bm}
\usepackage{hyperref}
\usepackage{cite}
\usepackage[utf8]{inputenc} 
\usepackage[T1]{fontenc}
\usepackage{graphicx}
\usepackage[font=small]{caption}
\usepackage{tikz}
\usetikzlibrary{positioning}
\usepackage{pgfplots}
\usepackage{epstopdf}
\usepackage[caption=false,font=footnotesize,labelformat=simple]{subfig}
\usepackage{multirow}
\usepackage{lipsum}
\usepackage{mathtools}

\long\def\comment#1{}

\newtheorem{lemma}{Lemma}

\input{mysymbol.sty}

\newcommand{\update}[1]{#1}

\begin{document}

\title{Enhanced Backpressure Routing Using Wireless Link Features
\thanks{Research was sponsored by the Army Research Office and was accomplished under Cooperative Agreement Number W911NF-19-2-0269. 
The views and conclusions contained in this document are those of the authors and should not be interpreted as representing the official policies, either expressed or implied, of the Army Research Office or the U.S. Government. 
The U.S. Government is authorized to reproduce and distribute reprints for Government purposes notwithstanding any copyright notation herein.
\newline
Emails: $^\star$\{zhongyuan.zhao, segarra\}@rice.edu, $^\ddag$\{gunjan.verma.civ, ananthram.swami.civ\}@army.mil}}

\author{
Zhongyuan Zhao$^\star$, Gunjan Verma$^\ddag$, Ananthram Swami$^\ddag$, and Santiago Segarra$^\star$\\
\textit{$^\star$Rice University, USA \hspace{10mm}  \hspace{2mm}  $^\ddag$US Army’s DEVCOM Army Research Laboratory, USA}
}

\maketitle

\begin{abstract}
Backpressure (BP) routing is a well-established framework for distributed routing and scheduling in wireless multi-hop networks.
However, the basic BP scheme suffers from poor end-to-end delay due to the drawbacks of slow startup, random walk, and the last packet problem. 
Biased BP with shortest path awareness can address the first two drawbacks, and sojourn time-based backlog metrics were proposed for the last packet problem. 
Furthermore, these BP variations require no additional signaling overhead in each time step compared to the basic BP. 
In this work, we further address three long-standing challenges associated with the aforementioned low-cost BP variations, including optimal scaling of the biases, bias maintenance under mobility, and incorporating sojourn time awareness into biased BP.
Our analysis and experimental results show that proper scaling of biases can be achieved with the help of common link features, which can effectively reduce end-to-end delay of BP by mitigating the random walk of packets under low-to-medium traffic, including the last packet scenario.
In addition, our low-overhead bias maintenance scheme is shown to be effective under mobility, and our bio-inspired sojourn time-aware backlog metric is demonstrated to be more efficient and effective for the last packet problem than existing approaches when incorporated into biased BP. 
\end{abstract}
\section{Introduction}\label{sec:intro}

Backpressure (BP) routing~\cite{tassiulas1992} is a well-established algorithm for distributed routing and scheduling in wireless multi-hop networks.
These networks have been widely used in military communications, disaster relief, and wireless sensor networks, and are envisioned to support emerging applications such as connected vehicles, robotic swarms, xG (device-to-device, wireless backhaul, and non-terrestrial coverage), Internet of Things, and machine-to-machine communications \cite{Lin06,sarkar2013ad,kott2016internet,akyildiz20206g,cisco2020,chen2021massive}.
The distributed nature of the BP algorithms  \cite{tassiulas1992,neely2005dynamic,georgiadis2006resource,jiao2015virtual,cui2016enhancing,gao2018bias,Alresaini2016bp,ji2012delay,athanasopoulou2012back,hai2018delay,Rai2017loop,yin2017improving,ying2010combining,Ying2012scheduling,Ryu2012timescale,zhao2023icassp} empowers wireless multi-hop networks to be self-organized without relying on infrastructure, and promotes scalability and robustness that are critical to many applications.
In BP, each node maintains a separate queue for packets to each destination (also denominated as commodity), routing decisions are made by selecting the commodity that maximizes the differential backlog between the two ends of each link, and data transmissions are activated on a set of non-interfering links via MaxWeight scheduling~\cite{tassiulas1992,joo2012local,zhao2021icassp}.
This mechanism uses congestion gradients to drive data packets towards their destinations through all possible routes, avoiding congestion and stabilizing the queues in the network for any flow rate within the network capacity region  \cite{tassiulas1992,neely2005dynamic,georgiadis2006resource} (this property is known as throughput optimality). 

However, classical BP routing is known to suffer from poor delay performance, especially under low-to-medium traffic loads   \cite{neely2005dynamic,georgiadis2006resource,jiao2015virtual,cui2016enhancing,gao2018bias}, 
exhibiting three drawbacks:
1)~\emph{slow startup}: when a flow starts, many packets have to be first backlogged to form stable queue backlog-based gradients, causing large initial end-to-end delay;
2)~\emph{random walk}: during BP scheduling, the fluctuations in queue backlogs drive packets towards random directions, causing unnecessarily long routes or loops;
and 
3)~the \emph{last packet problem} \cite{Alresaini2016bp,ji2012delay}, which refers to the phenomena that packets of a short-lived flow could remain enqueued in the network for a long time due to the absence of pressure.

To improve the latency, four types of BP variations have been developed: 
1)~Biased BP adds pre-defined queue-agnostic biases, e.g., (functions of) shortest path distance \cite{neely2005dynamic,georgiadis2006resource,jiao2015virtual,zhao2023icassp} to the backlog metric.
Shortest path-biased BP (SP-BP) can mitigate the drawbacks of slow startup and random walk, while maintaining the throughput optimality~\cite{neely2005dynamic,georgiadis2006resource,jiao2015virtual} at a low cost, i.e., a one-time communication overhead for bias computation.
2)~Delay-based BP replaces queue-length with delay metrics \cite{ji2012delay,cui2016enhancing,hai2018delay} for the backlog.
This approach addresses the last packet problem and maintains throughput optimality. 
3)~Impose restrictions on the routes~\cite{Rai2017loop,yin2017improving} or hop counts~\cite{ying2010combining} to prevent or reduce loops.
However, approaches (2) and (3) above may shrink the network capacity region.
4)~Use queue-dependent biases that aggregate the queue state information (QSI) of the local neighborhood (or global QSI) to improve the myopic BP decisions \cite{cui2016enhancing,gao2018bias} or use shadow queues \cite{athanasopoulou2012back,Alresaini2016bp} to dynamically increase the backpressure.
However, collecting neighborhood or global QSI at every time step increases communications overhead.  
Besides latency, other practical enhancements include extending BP to intermittently connected networks \cite{Ryu2012timescale,Alresaini2016bp} and uncertain network states \cite{Ying2012scheduling}.
In addition, most of the aforementioned approaches require careful parameter tuning, typically done via trial-and-error.

In this work, we aim to improve the SP-BP routing \cite{neely2005dynamic,georgiadis2006resource,zhao2023icassp} in the presence of the last packet problem and network mobility, while retaining its low computational and communication overhead and throughput optimality.
Specifically, we seek to address three longstanding challenges associated with SP-BP: 
1)~how to optimize the per-hop distance, in other words, the scaling of the shortest path bias; 2)~how to efficiently update the shortest path bias under node mobility, e.g., nodes moving, joining or leaving the network; and 3)~how to effectively incorporate sojourn time awareness into SP-BP in response to the last packet problem.

\vspace{1mm}
\noindent
{\bf Contribution.} The contributions of this paper are as follows:\\
1) We develop a principled approach to optimize the per-hop distance metric (bias scaling) based on features of wireless links, to mitigate the random walk behavior and the last packet problem in SP-BP. \\
2) We develop an efficient bias maintenance rule for node mobility based on local information exchange between 1-hop neighbors. \\
3) We further propose a new delay metric, \emph{expQ}, to prioritize old packets without having to track the sojourn time of individual packets, which improves the bandwidth efficiency of the network. \\ 
4) We demonstrate via numerical experiments that shortest path biases can address all three drawbacks of basic BP and are more effective than delay metrics \cite{ji2012delay,hai2018delay} in reducing end-to-end delay.

\section{(Biased) Backpressure Routing}
\label{sec:basic}

\noindent
{\bf System Model: }
We model a wireless multi-hop network as an undirected graph $\ccalG^{n}=(\ccalV, \ccalE)$, where $\ccalV$ is a set of nodes representing user devices in the network, and $\ccalE$ represents a set of links, where $e=(i,j)\in\ccalE$ for $i,j\in\ccalV$ represents that node $i$ and node $j$ can directly communicate.
$\ccalG^{n}$ is called a connectivity graph and assumed to be a connected graph, i.e., two arbitrary nodes in the network can always reach each other.
Notice that routing involves directed links, so we use ($\overrightarrow{i,j}$) to denote data packets being transmitted from node $i$ to node $j$ over link $(i,j)$.
There is a set of flows $\ccalF$ in the network, in which a flow $f=(i,c)\in\ccalF$, where $i\neq c$ and $i,c\in\ccalV$, describes the stream of packets from a source node $i$ to a destination node $c$, potentially through multiple links.
The medium  access control (MAC) of the wireless network is assumed to be time-slotted orthogonal multiple access. 
Each time slot $t$ contains a stage of decision making for routing and scheduling, followed by a second stage of data transmission.
Each node hosts multiple queues, one for each flow destined to node $c\in\ccalV$ (or packets of commodity $c$).
We use $\bbQ_{i}^{(c)}(t)$ to denote the queue of commodity $c$ at node $i$ at the beginning of time slot $t$, and $Q_{i}^{(c)}(t)$ for its queue length.

Matrix $\bbR\in\mathbb{Z}_{+}^{|\ccalE|\times T}$ collects the (stochastic) real-time link rates, of which an element $\bbR_{e,t}$ represents the number of packets that can be delivered over link $e$ in time  slot $t$.
The long term link rate of a link $e\in\ccalE$ is denoted by  $r_{e}=\mathbb{E}_{t\leq T}\left[\bbR_{e,t}\right]$, 
and $\bar{r}=\mathbb{E}_{e\in\ccalE,t\leq T}\left[\bbR_{e,t}\right]$ is the network-wide average link rate.

Under orthogonal multiple access, the conflict relationships between wireless links are captured by a \emph{conflict graph}, $\ccalG^c=(\ccalE,\ccalC)$, defined as follows: a vertex $e\in\ccalE$ represents a link in the network, and the presence of an undirected edge $(e_1, e_2)\in\ccalC$ means that simultaneous communications on links $e_1, e_2\in\ccalE$ cause interference.
We consider the scenario where all the users transmit at identical power levels with an omnidirectional antenna, which can be captured by the unit-disk interference model, in which two links conflict with each other if they share the same node or if any of their nodes are within a pre-defined distance.
For the rest of this paper, we assume the conflict graph $\mathcal{G}^{c}$ to be known, e.g., by each link monitoring the wireless channel \cite{zhao2022twc}.

\vspace{1mm}
\noindent
{\bf BP Algorithm: }
BP routing and scheduling consist of 4 steps.
In step 1, the optimal commodity $c_{ij}^{*}(t)$ on each {directed} link ($\overrightarrow{i,j}$) is selected as the one with the maximal backpressure, 
\begin{equation}\label{E:commodity}
    c_{ij}^{*}(t) =\argmax_{c\in\ccalV}\{ U_{i}^{(c)}(t) - U_{j}^{(c)}(t) \} \;,
\end{equation}
where $U_{i}^{(c)}(t)$ is the backlog metric, whose relationship with  queue lengths is discussed later. The backpressure of commodity $c$ on directed link ($\overrightarrow{i,j}$) is defined as  $U_{ij}^{(c)}(t) = U_{i}^{(c)}(t) - U_{j}^{(c)}(t)$.
In step 2, the maximum backpressure of ($\overrightarrow{i,j}$) is found as:
\begin{equation}\label{E:weight}
    w_{ij}(t) =\max\{ U_{i}^{(c_{ij}^*(t))}(t) - U_{j}^{(c_{ij}^*(t))}(t), 0 \}\;.
\end{equation}
In step 3, MaxWeight scheduling \cite{tassiulas1992} finds the schedule $\bbs (t)\in\{0,1\}^{|\ccalE|}$ to activate a set of \emph{non-conflicting links} achieving the maximum total utility, where the per-link utility is  $u_{ij}(t)=\bbR_{ij,t}\tilde{w}_{ij}(t)$, 
\begin{equation}\label{E:scheduling}
    \bbs (t) = \argmax_{\tilde{\bbs} (t)\in \ccalS } ~ \tilde{\bbs}(t)^\top  \left[\bbR_{*,t}\odot\tilde{\bbw}(t)\right] \;,
\end{equation}
in which $\ccalS$ denotes the set of all non-conflicting configurations,
vector $\bbR_{*,t}$ collects the real-time link rate of all links, vector $\tilde{\bbw}(t)=\left[ \tilde{w}_{ij}(t) | (i,j)\in\ccalE\right]$, where $\tilde{w}_{ij}=\max\{w_{ij}(t),w_{ji}(t)\}$, and the direction of the link selected by the $\max$ function will be recorded for step 4.
MaxWeight scheduling involves solving an NP-hard maximum weighted independent set (MWIS) problem \cite{joo2010complexity} on the conflict graph to find a set of non-conflicting links. 
In practice, \eqref{E:scheduling} can be solved approximately by distributed heuristics, such as local greedy scheduler (LGS)~\cite{joo2012local} and its GCN-based enhancement~\cite{zhao2022twc}.
In step 4,  all of the real-time link rate $\bbR_{ij,t}$ of a scheduled link is allocated to its optimal commodity $c_{ij}^{*}(t)$.
The final transmission and routing variables of commodity $c\in\ccalV$ on link ($\overrightarrow{i,j}$) is
\begin{equation}\label{E:quota}
    \mu_{ij}^{(c)}(t) = \begin{cases}
         \bbR_{ij,t}, & \text{if } c=c_{ij}^{*}(t), w_{ij}(t)>0, s_{ij}(t)=1, \\
         0, & \text{otherwise}.
    \end{cases}
\end{equation}

\vspace{1mm}
\noindent
{\bf (Biased) backlog metrics: }
The general form of the backlog metric, i.e., $U_{i}^{(c)}(t)$ in \eqref{E:commodity}, in low-cost BP schemes can be expressed as 
\begin{equation}\label{E:lowcost}
    {U}_{i}^{(c)}(t) = g\left(Q_{i}^{(c)}(t)\right) + B_{i}^{(c)}  \;,
\end{equation}
where $g(\cdot)$ is a function of QSI, and $ B_{i}^{(c)} $ is a queue-agnostic bias.
In queue length-based BP \cite{tassiulas1992,neely2005dynamic,georgiadis2006resource,jiao2015virtual,zhao2023icassp}, $g(Q_{i}^{(c)}(t))=Q_{i}^{(c)}(t)$, whereas in delay-based BP, $g(Q_{i}^{(c)}(t))$ can be the sojourn time of the head-of-line (HOL) packet  \cite{ji2012delay} or the entire queue, i.e., sojourn time backlog (SJB) \cite{hai2018delay}.
In unbiased BP, such as the basic and delay-based BP, $B_{i}^{(c)}=0$.
In biased BP \cite{neely2005dynamic,georgiadis2006resource,jiao2015virtual,zhao2023icassp}, $B_{i}^{(c)}\geq0$ is defined based on the shortest path distance between node $i$ and destination $c$.
The set of biases is denoted as $\ccalB=\{B_{i}^{(c)}|i,c\in\ccalV\}$.

The shortest path distances can be computed by distributed algorithms for single source shortest path (SSSP) or all pairs shortest path (APSP) on the initialization of a node or network, and updated periodically to capture the evolving network topology.
On weighted graphs, the distributed SSSP with the Bellman-Ford algorithm \cite{bellman1958routing,ford1956network} and state-of-the-art APSP~\cite{bernstein2019distributed} both take $\ccalO(|\ccalV|)$ rounds. 
On unweighted graphs, the distributed SSSP and APSP take $\ccalO(D)$ and $\ccalO(|\ccalV|)$  rounds, respectively, where $D$ is the diameter of the unweighted $\ccalG^{n}$.

\section{Improvements to Biased Backpressure}
\label{sec:solution}

We focus on SP-BP routing~\cite{neely2005dynamic,zhao2023icassp}, where the length of every edge $e\in \ccalE$ in the shortest path computation, {namely the \emph{per-hop distance}, denoted as $\delta_e$}, is scaled by a constant.
For example, $\delta_{e}=K$ for all $e\in \ccalE$ in enhanced dynamic routing (EDR) \cite{neely2005dynamic} 
and $\delta_{e} = R/(x_{e}r_{e})$ in \cite{zhao2023icassp}, where $x_e$ is the link duty cycle estimated by a graph convolutional network (GCN), and $R$ is the scaling constant.
Our methods are to i) optimize the scaling constant in the per-hop distance ($K$ in \cite{neely2005dynamic} and $R$ in \cite{zhao2023icassp}), ii) efficiently update the biases under node mobility, and iii) prioritize older packets over newer ones.

\vspace{1mm}
\noindent
{\bf Optimal scaling of per-hop distance.}
To the best of our knowledge, the parameter(s) for the scaling of the per-hop distance in biased BP are often selected via trial-and-error, whereas our goal is to provide a formal method of optimization for such scaling parameter(s).

\begin{figure}[t]
\centering
\subfloat[Time $t$]{
    \includegraphics[height=1.8in]{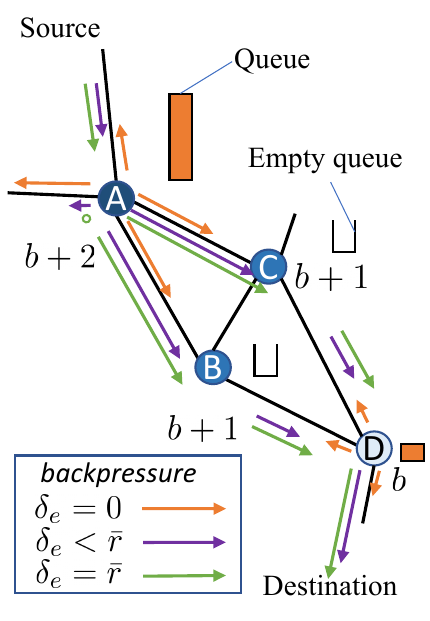}
    \label{fig:example:t0}\vspace{-0.1in}
}
\subfloat[Time $t+1$]{
    \includegraphics[height=1.8in]{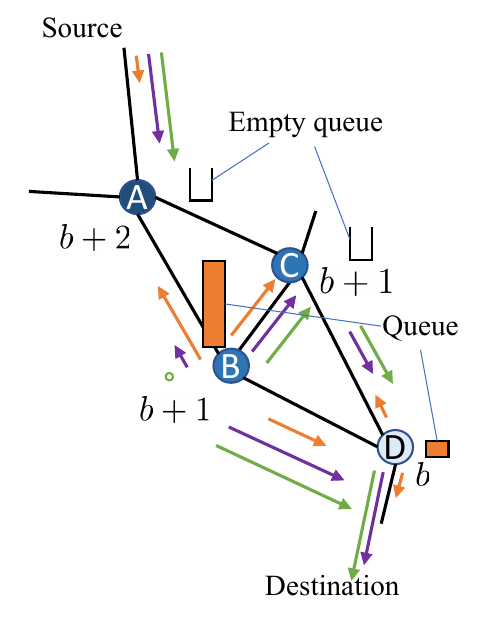}
    \label{fig:example:t1}\vspace{-0.1in}
}
    \caption{{\small The queue states at four nodes (with hop distance to destination marked) in an exemplary case of backpressure routing with a single commodity, at time (a) $t$ and (b) $t+1$. An arrow indicates the magnitude and direction of pressure on a link, its color encodes the choice of per-hop distance. All links have an identical rate of $\bar{r}$.
    }
    } 
 \label{fig:example}    
\end{figure}

A key observation is that the scaling of the per-hop distance in biased BP influences the degree of random walk of packets and subsequently, the end-to-end delay, under low to medium traffic loads, including the last packets situation.
Consider an exemplary situation of last packets in BP routing with only one commodity (flow), for which the queue states of four nodes in two consecutive time steps are illustrated in Fig.~\ref{fig:example}. 
The hop distances of nodes $A$, $B$, $C$, $D$ to the destination are $b+2$, $b+1$, $b+1$, $b$, respectively. 
Every link has a constant link rate of $\bar{r}$.
At time $t$, $Q_{A}^{(c)}(t)=\bar{r}$ and $Q_{B}^{(c)}(t)=Q_{C}^{(c)}(t)=0$, the link pressures point from $A$ towards $B$ and $C$. Assume that at time $t+1$, all the packets on $A$ moved to $B$.
Different choices of $\delta_{e}$ would lead to different routing decisions at $t+1$.

For unbiased BP, i.e., $\delta_{e}=0$, the backpressures (orange arrows) are pointed from $B$ towards $A$, $C$, and $D$, and $U^{(c)}_{BA}(t+1) = U^{(c)}_{BC}(t+1) = \bar{r} > U^{(c)}_{BD}(t+1)$. 
At the end of time slot $t+1$, the packets on $B$ will move to either $A$ or $C$, packets on $D$ may also move back to $C$, causing an unwanted meandering of the packets.
For biased BP, i.e., $\delta_{e}>0$, the backpressures (magenta and green arrows) at time $t+1$ originating from $B$ are skewed towards the destination, $U^{(c)}_{BD}(t+1) > U^{(c)}_{BC}(t+1) =\bar{r} > U^{(c)}_{BA}(t+1) $.
At the end of time slot $t+1$, packets in $B$ will move forward to $D$.
In particular, when $\delta_e=\bar{r}$, the backpressure (i.e., green arrows) $U^{(c)}_{BA}(t+1)=0$.

The previous example can be formally described as follows: 
in a wireless multi-hop network with homogeneous link rate, i.e., all links have a link rate of $\bar{r}$, 
there are two links $(i,j)$ and $(j,k)$, where nodes $j, k$ are on the shortest path from node $i$ to node $c$; 
consider a congestion-free last packets (CFLP) scenario: 
at time $t$, the last packets of commodity $c$ reside on node $i$, and 
link $(i,j)$ is congestion-free, i.e., $ 0< Q_{i}^{(c)}(t)=q\leq \bar{r}, Q_{j}^{(c)}(t)=0 $ and no external packets of commodity $c$ will arrive at nodes $i,j,k$ from the rest of the network or users. 
If link $(i,j)$ is scheduled at time $t$, such that $ Q_{i}^{(c)}(t+1)=0, Q_{j}^{(c)}(t+1)=q\leq \bar{r} $, we have the following.
\begin{lemma}\label{l:back}
In SP-BP routing under the CFLP scenario, the per-hop distance $\delta_e$ should be greater than or equal to the homogeneous link rate,  $\delta_e\geq\bar{r}$, to avoid the immediate reversal of the direction of backpressure for commodity $c$ on the scheduled link $(\overrightarrow{i,j})$ after the transmission. 
\end{lemma}
\vspace{-0.1in}
\begin{proof} 
In queue length-based SP-BP, the backpressure on link ($i,j$) is  $U_{ij}^{(c)}(t) = Q_{ij}^{(c)}(t) + (B_{i}^{(c)}-B_{j}^{(c)}) = q+\delta_{e}$. 
Since link $(i,j)$ is scheduled at time $t$, we have that $U_{ij}^{(c)}(t+1)=-q+\delta_{e}$. 
To avoid the direction of the backpressure on link $(i,j)$ being reversed at time $t+1$, i.e., $ U_{ij}^{(c)}(t+1)<0 $, we need to set $\delta_{e}\geq\bar{r}$.  
\end{proof}

In the described CFLP scenario, Lemma~\ref{l:back} ensures that the backpressure algorithm does not oscillate as the last packets travel through the shortest path when $\delta_e\geq\bar{r}$.
On the other hand, a smaller value of $\delta_e$ is preferable for path finding and congestion prevention.
Indeed, an extremely large SP bias scaling $\delta_e$ would force every packet through the shortest path, thus hindering any information from the queue lengths and defeating the original purpose of BP routing.
Combining these two observations, \emph{we advocate for setting the minimal per-hop distance as the average link rate}, i.e., $\underset{e\in\ccalE}{\min}\;\delta_e\vcentcolon=\bar{r}$. In particular, for EDR in~\cite{neely2005dynamic}, this implies setting $\delta_e = K \vcentcolon=\bar{r}$.

Although the above discussion is based on the CFLP setting with all link rates equal to the average $\bar{r}$, in Section~\ref{sec:results} we demonstrate the experimental optimality of the above choice under heterogeneous link rates and general traffic settings.

\vspace{1mm}
\noindent
{\bf Bias maintenance.} 
Due to the mobility and distributed nature of networks, bias maintenance requires frequent or periodical SSSP and/or APSP computation.
Although the biases $\ccalB$ can be re-used for many time slots to match the slowly changing topology, the overhead of bias maintenance is still high for large networks.

To address this issue, we propose a neighborhood update rule for bias maintenance at node $i\in\ccalV$, whenever one or more of its incidental links are established or destroyed 
\begin{equation}\label{E:neighbor}
    B_{i}^{(c)}(t+1) = \begin{cases}
    \underset{j\in\ccalN(i)}{\min} \left[B_{j}^{(c)}(t) + \delta_{ij}(t) \right], \;& i \neq c \\
    0, & i = c
    \end{cases}\;.
\end{equation}
where $\delta_{ij}(t)$ is the per-hop distance between neighboring nodes $i$ and $j$.
This rule allows the shortest path bias to be updated within $\ccalO(D)$ steps of local message exchange.

\begin{figure*}[t]
\centering
 \vspace{-0.2in}
\subfloat[]{
    \includegraphics[width=0.31\linewidth]{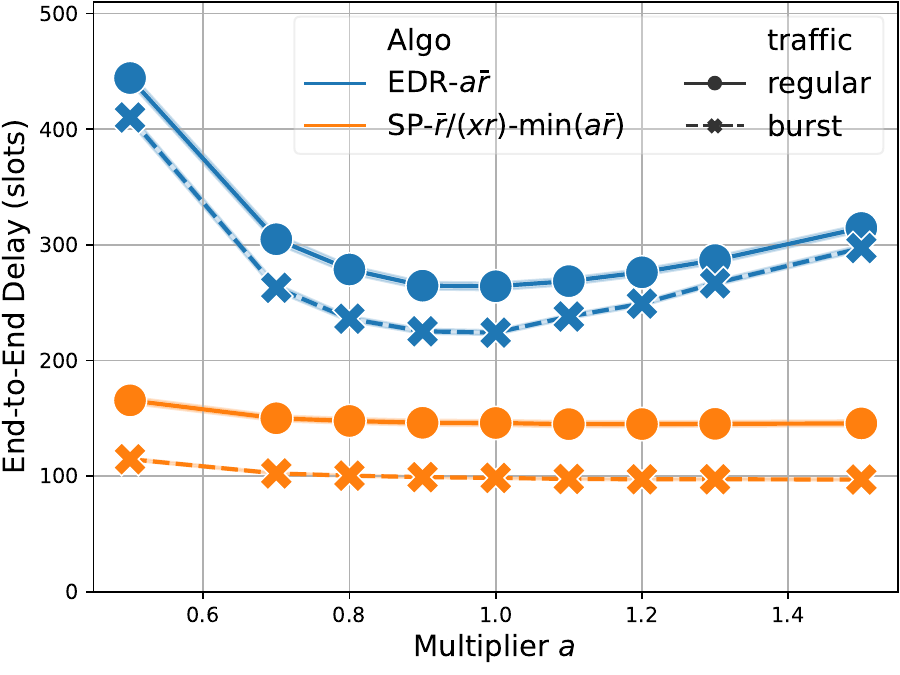}
    \label{fig:regular:mpx}\vspace{-0.1in}
}
\subfloat[]{
    \includegraphics[width=0.31\linewidth]{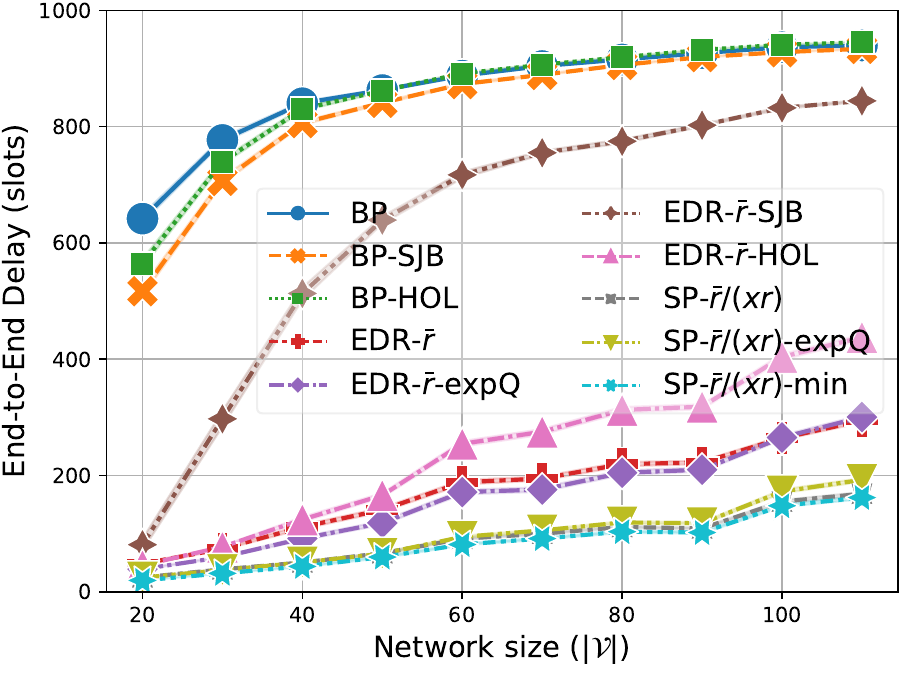}
    \label{fig:regular:delay}\vspace{-0.1in}
}
\subfloat[]{
    \includegraphics[width=0.31\linewidth]{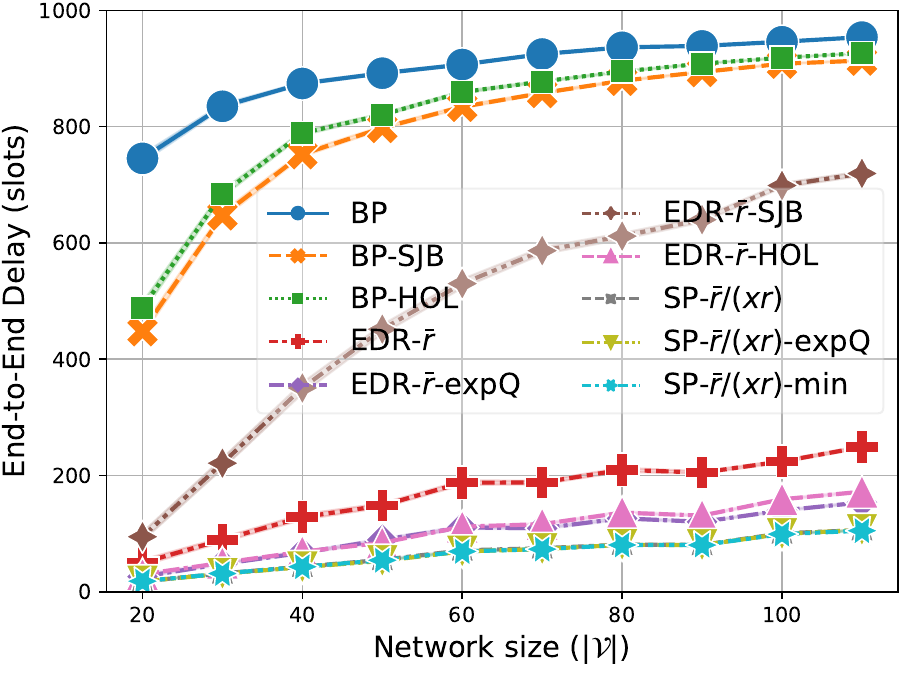}
    \label{fig:burst:delay}\vspace{-0.1in}
}
    \vspace{-0.1in}
    \caption{{\small End-to-end delay of backpressure routing algorithms, with unit-disk interference model, long-term link rates $r_{e} \sim \mathbb{U}(10,42)$, simulation time steps $T=1000$, and $100$ test instances per point (10 random networks $\times$ 10 realizations of random source-destination pairs and random link rates). 
    (a)~Delay vs. varying minimal per-hop distance multiplier $\min(\delta_{e})=a\bar{r}$ for SP-BP on networks of $100$ nodes. 
    (b)~Delay under streaming traffic, and 
    (c)~Delay under bursty traffic, v.s. network size (of 20-110 nodes).   
    Zoom in to view the error band ($95\%$ confidence interval) in light colors.
    }
    } 
 \label{fig:regular}   
 \label{fig:burst}
 \vspace{-0.1in}
\end{figure*}

\vspace{1mm}
\noindent
{\bf Sojourn time-aware backlog (expQ).}
To prioritize old packets, we further introduce a delay-aware QSI function $g(\cdot)$ in \eqref{E:lowcost}: 
\begin{equation}\label{E:expQ}
\begin{aligned}
    g\!\left(Q_{i}^{(c)}(t+1)\!\right)\!=&(1+\epsilon)g\!\left(Q_{i}^{(c)}(t)\!\right)\!\left[1-\frac{\Delta Q^{(c)}_{i,Tx}(t+1)}{Q_{i}^{(c)}(t)}\right] \\
    & + \Delta Q^{(c)}_{i,Rx}(t+1)
\end{aligned}
\end{equation}
where $\epsilon\geq 0$ is a small constant parameter, $  \Delta Q^{(c)}_{i,Tx}(t) $  and $ \Delta Q^{(c)}_{i,Rx}(t) $ are the numbers of packets that left and arrived at the queue $Q^{(c)}_{i}$ at node $i$ at time $t$.
For $\epsilon=0$, $ g(Q_{i}^{(c)}(t))=Q_{i}^{(c)}(t) $.

The $g(\cdot)$ in \eqref{E:expQ} is named as expQ, as it will increase exponentially over time if commodity $c$ at node $i$ is not scheduled.
Compared to the backpressure metric based on sojourn time backlog \cite{hai2018delay} and HOL sojourn time \cite{ji2012delay}, \update{expQ can increase the bandwidth utilization efficiency, since it does not require tracking the sojourn time of individual packets.} 

\vspace{1mm}
\noindent
{\bf Throughput optimality.}
Queue length-based biased BP schemes with queue-agnostic biases automatically inherit the throughput optimality of basic BP as long as the biases are non-negative and queue-agnostic, according to the proof in \cite{neely2005dynamic}.

\section{Numerical experiments}
\label{sec:results}

We evaluate various low-cost BP routing algorithms on simulated wireless multi-hop networks, generated by a 2D point process.
The simulations are configured similar to those in \cite{zhao2023icassp}, with only three differences:
a) Each test instance contains a number (uniformly chosen between $\lfloor 0.30 |\ccalV|\rfloor$ and $ \lceil 0.50|\ccalV|\rceil$) of random flows.
b) Only a unit-disk interference model is used.
c) Two types of traffic are employed: Under streaming traffic, {the arrival rate of a flow} $\lambda(f) \sim \mathbb{U}(0.2,1.0)$ for all time slots. 
Under bursty traffic, {the flow arrival rate} $\lambda(f) \sim \mathbb{U}(2.0,10.0)$ for $t<30$, and no packet arrivals for $t\geq30$.  
The end-to-end delay of a test BP scheme is collected by tracking the time each packet arrived at the source node $t_0$ and the destination node (departure time) $t_1$.
To be conservative, we treat the delay of an undelivered packet as $T-t_0$.

The tested BP algorithms (acronyms in parentheses identify legends in Fig. 2) include 1) unbiased BP: vanilla BP (BP), delay-based unbiased  BP with SJB \cite{hai2018delay} (BP-SJB) and HOL \cite{ji2012delay} (BP-HOL); 2) queue-length based biased BP: EDR \cite{neely2005dynamic} with per-hop distance $\delta_{e}=\bar{r}$ (EDR-$\bar{r}$), and GCN-based delay-aware shortest path bias \cite{zhao2023icassp} (SP-$\bar{r}/(xr)$) with per-hop distance $\delta_{e}=\bar{r}/(x_{e}r_{e})$, and its scaled version (SP-$\bar{r}/(xr)$-min) with per-hop distance $\tilde{\delta}_{e}={\delta}_{e}\bar{r}/(\min_{e\in\ccalE}{\delta}_{e})$; and
3) delay-based biased BP: combination of BP-SJB and EDR-$\bar{r}$ (EDR-$\bar{r}$-SJB), combination of BP-HOL and EDR-$\bar{r}$ (EDR-$\bar{r}$-HOL), and the two biased BP schemes based on expQ (EDR-$\bar{r}$-expQ and SP-$\bar{r}/(xr)$-expQ) where $\epsilon=0.01$ in~\eqref{E:expQ}. 
\update{The end-to-end delay of the tested BP schemes are illustrated in Figs.~\ref{fig:regular}, where each point in the plot is obtained by averaging 100 test instances, specifically, on 10 random networks of a particular size, each with 10 realizations of random source-destination pairs and random link rates}. 

\vspace{1mm}
\noindent
{\bf Optimal per-hop distance adjustment.}
We test EDR-$\delta$ and SP-$\bar{r}/(xr)$-min with per-hop distance adjustment $\underset{e\in\ccalE}{\min}\;\delta_e\vcentcolon=a\bar{r}$, for $a\in\left[0.5,1.5\right]$, on random networks of 100 nodes. 
The results in Fig.~\ref{fig:regular:mpx} show that $a=1.0$ is indeed the optimal setting for EDR-$\delta$ and near-optimal setting for SP-$\bar{r}/(xr)$-min, under streaming and bursty traffics.
This validates the optimality of adjustment rule $\underset{e\in\ccalE}{\min}\;\delta_e\vcentcolon=\bar{r}$ advocated in Section~\ref{sec:solution}.
Notice that the settings tested in Fig.~\ref{fig:regular:mpx} go beyond the simplified setting of CFLP with constant rates treated in Section~\ref{sec:solution}.
However, the proposed scaling is still empirically optimal in these broader scenarios.
Since SP-$\bar{r}/(xr)$-min can tolerate a wide range of $a$, to keep its distributed execution, it can be implemented based on statistical information of per-hop distances rather than their global minimal.
{The different sensitivities of the end-to-end delay to the scaling of per-hop distance under the two tested SP-BP schemes, as shown in Fig.~\ref{fig:regular:mpx}, reveal that the shortest path based on link features \cite{zhao2023icassp} not only achieves better performance but is also more robust to the scaling choice.}

\vspace{1mm}
\noindent
{\bf Streaming traffic.}
The average end-to-end delays (in time slots) of the 10 routing algorithms under streaming traffic, as a function of the network size are presented in Fig.~\ref{fig:regular:delay}.
Most BP variations outperform the vanilla BP.
However, the benefit of delay-based unbiased BP (BP-SJB and BP-HOL) diminishes on larger networks, as unbiased BP schemes still suffer from the drawbacks of slow startup and random walk, leading to very low delivery rates.

Biased BP based on queue length (EDR-$\bar{r}$, SP-$\bar{r}/(xr)$, and SP-$\bar{r}/(xr)$-min) can significantly reduce the end-to-end delay, as the pre-defined distance gradients lead to fast startup and mitigated random walk. 
SP-$\bar{r}/(xr)$ can reduce the delay of EDR-$\bar{r}$ by nearly half, which is consistent with the results in \cite{zhao2023icassp}.
Notice that SP-$\bar{r}/(xr)$-min only slightly improves SP-$\bar{r}/(xr)$.
Delay metrics such as SJB and HOL do not directly benefit biased BP with our per-hop distance adjustment, which is based on queue length.
As shown in Fig.~\ref{fig:regular:delay},
EDR-$\bar{r}$-SJB and EDR-$\bar{r}$-HOL both significantly underperform EDR-$\bar{r}$,
whereas expQ only slightly improves EDR-$\bar{r}$ for smaller networks ($|\ccalV|\leq 90$), while slightly degrading SP-$\bar{r}/(xr)$.

\vspace{1mm}
\noindent
{\bf Bursty traffic.}
The end-to-end delays of the 10 BP algorithms under bursty traffic are presented in Fig.~\ref{fig:burst:delay}, which shows their effectiveness in the last packet problem. 
Compared to streaming traffic, every BP scheme performs better under bursty traffic, except vanilla BP.
EDR-$\bar{r}$ is substantially improved by both HOL and expQ, 
where the latter slightly outperforms the former. 
For SP-$\bar{r}/(xr)$, the benefits of per-hop distance adjustment and expQ are negligible since SP-$\bar{r}/(xr)$ already reaches $100\%$ delivery rate. 
This shows that the last packet problem can be significantly mitigated by our proposed per-hop distance scaling, as EDR-$\bar{r}$ can achieve a high delivery rate of $93.3\%$, and further improved by combining with HOL and expQ with delivery rate $99.7\%$ and $99.9\%$, respectively. 
Compared to HOL, expQ is not degraded under streaming traffic and does not require tracking the sojourn time of individual packets.

\begin{table}[thbp]
	\renewcommand{\arraystretch}{1.2}
	\caption{End-to-end delay on random networks of $100$ nodes, $T=1000$ under unit-disk interference model and node mobility. 
	} 
	\label{tab:mobility}
	\centering
    \scriptsize
	\begin{tabular}{|l|l|l|l|l|}
        \hline
	    & \multicolumn{2}{|c|}{EDR-$\bar{r}$} & \multicolumn{2}{|c|}{SP-$\bar{r}/(xr)$}  \\ \hline
        Bias update & Delay (std.)  & Delivery (std.) & Delay (std.) & Delivery (std.) \\ \hline
        Ideal & 278.3 (106.1) & 81.2\% (8.7\%) & \textbf{155.2} (80.9) & \textbf{90.8\%} (6.1\%) \\ \hline
        Neighbor & 331.8 (103.8) & 75.7\% (9.1\%) & \textbf{282.5} (87.5) & \textbf{78.5\%} (7.8\%) \\ \hline
	\end{tabular}
\end{table}

\vspace{1mm}
\noindent
{\bf Node mobility.}
Lastly, we run 100 test instances on random networks with 100 nodes and node mobility, in which for every 100 time steps, $ 10 $ random nodes take a random step ($\Delta \bbx \sim \mathrm{N}^2(0,0.1) \in \reals^2 $) on the 2D plane, while keeping the network connected. 
The long-term link rate of a newly created link $e$ is set as $r_{e} \sim \mathbb{U}(10,42)$.
We compare the delay and delivery rate of our neighborhood bias update rule in \eqref{E:neighbor} with those of ideal (instantaneous) SSSP in Table~\ref{tab:mobility}.
Under the practical neighborhood update rule, SP-$\bar{r}/(xr)$ suffers a larger loss from the ideal (but unpractical) SSSP than EDR-$\bar{r}$, i.e., the delivery rate of SP-$\bar{r}/(xr)$  dropped by $12\%$ compared to a drop of $5.6\% $ with EDR-$\bar{r}$.
Despite larger delays and lower delivery rates, EDR-$\bar{r}$ suffers less under node mobility likely due to its simplicity.

\section{Conclusions}
\label{sec:conclusions}
In this paper, we improve the latency and practicality of shortest path-based Backpressure routing by introducing three components: optimal bias scaling, low-overhead bias maintenance, and a delay-aware backlog metric. 
These enhancements are demonstrated to be effective in wireless multi-hop networks with relatively low mobility. 
In the future, it is worthwhile to investigate and further improve the performance of our approach in scenarios with varying network mobility, mixed streaming and bursty traffics, as well as uncertainties and shifts in link features and their global minimal values. 

\bibliographystyle{ieeetr}
{
\bibliography{strings,refs}
}

\end{document}